\documentclass[11pt]{evn}
\usepackage{graphicx}
\usepackage{epsf}
\usepackage[english]{babel}
\long\def\maintitle#1{{\vskip 20mm \begin{center}\section*{#1}\end{center}\nopagebreak[4]}}

\long\def\author#1{{\begin{center}\normalsize{\bf#1}\end{center}\vskip-1em\index{#1}}\nopagebreak[4]}
\long\def\address#1{{\begin{center}\small\noindent#1\end{center}\vskip-8mm}\nopagebreak[4]}

\begin{document}
\noindent\mbox{\small The 13$^{th}$ EVN Symposium \& Users Meeting Proceedings, 2016}

\maintitle{APSYNSIM: An Interactive Tool To Learn Interferometry}

\author{I.~Marti-Vidal$^{1}$}

\address{$^{2}$Onsala Space Observatory, Chalmers University (Sweden)}

\begin{abstract}
The {\em AP}erture {\em SYN}thesis {\em SIM}ulator is a simple interactive tool to help the students visualize and understand the basics of the Aperture Synthesis technique, applied to astronomical interferometers. The users can load many different interferometers and source models (and also create their own), change the observing parameters (e.g., source coordinates, observing wavelength, antenna location, integration time, etc.), and even deconvolve the interferometric images and corrupt the data with gain errors (amplitude and phase). The program is fully interactive and all the figures are updated in real time. APSYNSIM has already been used in several interferometry schools and has got very positive feedback from the students.
\end{abstract}

\section{Introduction}

{\em APSYNSIM} ({\em AP}erture {\em SYN}thesis {\em SIM}ulator) is a Python-based interactive tool to simulate simple astronomical interferometric observations. This tool is useful to help the students visualize and understand the basics of the Aperture Synthesis technique (e.g., \cite{1}). APSYNSIM uses the Numpy and Matplotlib (open source) packages \cite{2}. The {\em APSYNSIM} code is also open (distributed under the GPL version 3) and can be downloaded from:\\

\texttt{https://launchpad.net/apsynsim}\\

\begin{figure}[ht!]
\centering
\includegraphics[width=10cm]{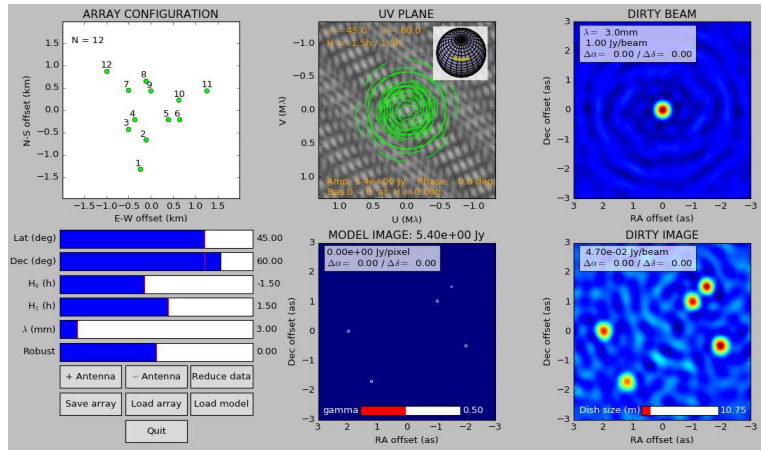}
\caption{Main APSYNSIM Window. Five compact sources are being observed with a Golay array of 12 elements.}
\label{FIG1}
\end{figure}

Binary pre-compiled packages for MS Windows and Mac OS can also be downloaded. Any image (JPG, BMP, PNG, etc.) can be used as a source. The image size (in arc seconds) and the flux-density scale (in Jy per pixel) can be set via an ascii file. Other simple structures can be added to the source model, like Gaussians, disks, rings, etc. The user can also load many different interferometer arrays (and define his/her own arrays, as simple ascii configuration files). It is possible to change the observing conditions in real time (e.g., antenna positions, source coordinates, observing wavelength, integration time, visibility weighting, etc.). Once the observing (and imaging) parameters are all set, it is possible to load a deconvolution Graphical User Interface (GUI) to deconvolve the interferometric images using the CLEAN algorithm. We show in Fig.\,\ref{FIG1} the main window of APSYNSIM with an example observation. The array being used in this case is a Golay array of 12 elements (Golay arrays have the minimum baseline redundancy).

\section{More advanced simulations}

\subsection{Data corruption}

\begin{figure}[ht!]
\centering
\includegraphics[width=10cm]{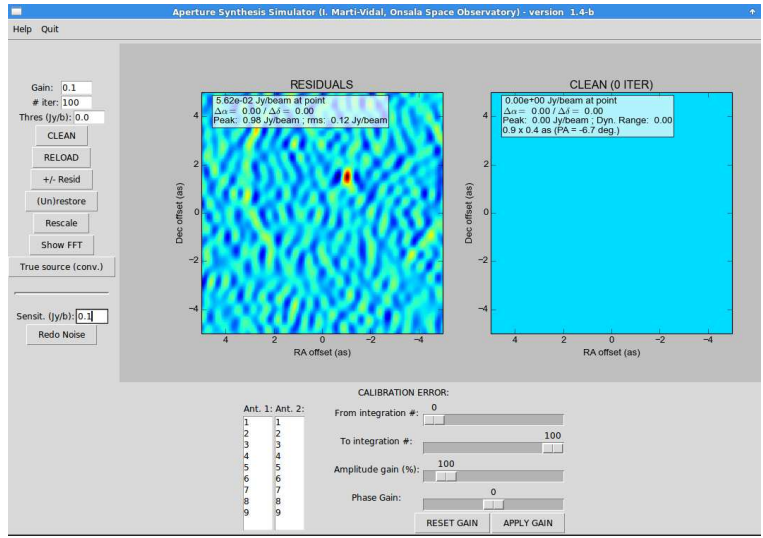}
\caption{Deconvolution GUI. Gain corruption and thermal noise can be added to the data through this GUI.}
\label{FIG2}
\end{figure}

The deconvolution GUI (Fig. \ref{FIG2}) offers the possibility of corrupting the data with antenna-based and/or baseline-based gain errors (in both amplitude and phase). Any phase offset (and/or amplitude gain ratio) can be applied to a given baseline or to all baselines of a given antenna. It is also possible to add thermal noise to the visibilities, by setting the desired image root-mean-square (rms), assuming natural weighting. This way, the effect of different visibility weightings on sensitivity-limited observations can be studied.

A very good exercise is to open different deconvolution GUIs and load different noises and/or corruption gains on each GUI, in order to compare, separately, different corruption effects during the imaging. This useful trick will work as long as the user opens {\em all} the GUIs to be used {\em before} corrupting the data (since the corrupted data are sent to the Main Window, and those data will be loaded by any GUI that is open afterwards), and as long as the user does not press the ``RELOAD'' button in the GUIs during the imaging (Fig.\,\ref{FIG2}), since this tells the GUI to re-read the visibilities from the Main Window.

\subsection{Primary beam}

Primary-beam effects are added to the observations by setting the antenna diameters. It is possible to define heterogeneous observations, where two interferometers with different antenna sizes observe the same source. This is, for instance, the case of the ALMA Main Array (12\,m antennas) and its complementary Atacama Compact Array (ACA, 7\,m antennas). Using heterogeneous subarraying has implications on the achievable dynamic range if extended sources are observed. The user can set the relative subarray weights in the imaging, to check how the sensitivity to the different spatial scales depends on the weight between a compact array (i.e., ACA) and more extended baselines (i.e., ALMA).

\subsection{Deconvolution in Fourier Space}

The deconvolution GUI also offers the possibility of checking the Fourier transforms (FTs) of the CLEAN model (both restored and unrestored), the source structure (the true structure and the one convolved with the CLEAN beam) and the image of residuals. This way, several deconvolution artifacts can be seen in Fourier space. For instance, in the case of gain corruption, the FT of the residuals will ``iluminate'' the tracks of the baselines with a wrong calibration; effects of missing flux-density, however, will be seen as large unmodelled residuals in the shortest baselines.

\section{Example experiments}

The software documentation comes with more detailed usage instructions and a set of example exercises. Here we summarize a subset of exercises that have been particularly useful for the students, so far:

\begin{itemize}

\item {\bf The two-element interferometer: } study the Point Spread Function (PSF) for snapshots and after applying some Earth Rotation Synthesis. Check the dependence on latitude, source declination, and position angle of the baseline.

\item {\bf East-West arrays: } study the UV coverage and PSF as a function of latitude, source declination, and hour-angle coverage. Compare cases of maximum baseline redundancy (i.e., a constant distance between neighboring antennas) with cases of minimum redundancy (i.e., Golomb rulers).

\item {\bf Missing flux density: } image a source with an extended component and drag one of the antennas to a position very close to another antenna. With a good hour-angle coverage, the extended emission will be more or less prominent in the dirty image, depending on the shortest baseline (i.e., depending on how close the antenna is dragged close to another antenna). Observing at longer wavelengths with a given antenna configuration will also make the dirty image more sensitive to the extended structure. CLEAN can later be applied, to check how much flux-density is recovered from the extended structure and to check the typical image artifacts related to the missing-flux deconvolution bias.

\item {\bf Assessing gain corruption: } corrupting gains can be applied before starting the deconvolution. Once CLEAN has been executed during some iterations, the user can rescale the color scale of the image residuals (pressing the ``Rescale'' button, Fig.\,\ref{FIG2}) to better see the typical symmetric/antisymmetric artifacts related to amplitude/phase gain corruption factors. Fourier-inverting the residuals (button ``Show FFT'') will show the user which baselines are producing such residual artifacts. The user can then compare these plots with the actual array distribution in the main APSYNSIM window.

\end{itemize}


\begin{thebibliography}{5}
\bibitem{1}
{\em Thompson~A.~R., Moran~J.~M., Swenson~G.~W.} Interferometry and Synthesis in Radio Astronomy // New York : Wiley. ~-- 2001. 

\bibitem{2}
{\em Hunter~J.~D.} Matplotlib: A 2D graphics environment // Computing In Science \& Engineering.~-- 2007.~-- Vol.~9. ~-- Num.~3.~-- P.~90--95.


\end{thebibliography}
\end{document}